\documentstyle[prl,multicol,aps,epic,eepic]{revtex}
\begin{document}
\draft
\preprint{}
\title{Two-mode heterodyne phase detection}
\author{G. M. D'Ariano\cite{dar} and M. F. Sacchi\cite{sac}}
\address{Dipartimento di Fisica ``Alessandro Volta'', Universit\`a degli
         Studi di Pavia, Via A. Bassi 6, I--27100 Pavia, Italy}
\maketitle
\begin{abstract}
We present an experimental scheme that achieves ideal phase detection 
on a two-mode field. The two modes $a$ and $b$ are the signal and image 
band modes of an heterodyne detector, with the field approaching an 
eigenstate of the photocurrent $\hat{Z}=a+b^{\dag}$. The field is obtained 
by means of a high-gain phase-insensitive amplifier followed by a 
high-transmissivity beam-splitter with a strong local oscillator at the 
frequency of one of the two modes.
\end{abstract}
\pacs{PACS number(s): 03.65.Bz, 42.50.Dv} 
\begin{multicols}{2}
The quantum-mechanical measurement of the phase of the radiation field is
the essential problem of high sensitive interferometry, and has
received much attention in quantum optics \cite{rev1,rev2}.  Most of
the work has been devoted to measurements on a single-mode
electromagnetic field, where the measurement cannot be achieved
exactly, even in principle, due to the lack of a unique self-adjoint
operator \cite{pom}.  \par It can be readily recognized that the
absence of a proper self-adjoint operator in the one-mode case is
mainly due to the semiboundedness of the spectrum of the number
operator \cite{shsh,ban}, which is canonically conjugated to the phase
in the sense of a Fourier-transform pair \cite{shap}. This observation
discloses the route toward an exact phase measurement in terms of
two-mode fields, where a phase-difference operator becomes conjugated
to an unbounded number-difference operator \cite{luis}. Moreover, as
already noticed in Ref.~\cite{shwa}, a two mode field corresponds to a
complex photocurrent $\hat Z$ such that $[\hat{Z},\hat{Z}^{\dag}]=0$,
with a self-adjoint phase operator $\hat{\phi}=\arg(\hat{Z})$ that can
concretely be measured.  Despite its promising possibilities, not much
work has been devoted to the two-mode phase detection, and attention
has been focused mostly on the algebraic structure the photocurrents
(see Refs. \cite{ban,shap,luis} and references therein). Only in
Ref. \cite{shwa} a concrete experimental set-up has been devised,
based on unconventional field heterodyning with the signal and
image-band modes both nonvacuum.  \par Here in this letter, following
the route opened by Ref.~\cite{shwa}, we study the eigenstates of the
heterodyne photocurrent $\hat{Z}$ and provide an experimental scheme
that approaches them. We then analyze the measurement of the two-mode
phase $\hat{\phi}=\arg(\hat{Z})$ showing that the ideal sensitivity
limit $\delta \phi =1/\overline{n}$ can be achieved for large mean
number of photons $\overline{n}$.  \par It has been proved by Yuen and
Shapiro \cite{yuen} that the output photocurrent $\hat Z$ of a
heterodyne detector (for unit quantum efficiency, and in the limit of
strong local oscillator and vanishing beam splitter reflectivity) is
just the operator $\hat{Z}=a+b^{\dag}$, where $a$ denotes (the
annihilator~of) the signal mode, and $b$ the image-band
mode. In ordinary heterodyning the image-band mode $b$ is vacuum, and
is responsible for the additional 3dB noise. Here, similarly to
Ref.~\cite{shwa}, we use the heterodyne detector in an unconventional
way, namely with a nonvacuum $b$ mode, and look for field states which
are eigenvectors of the current $\hat Z$.  \par \noindent It is easy
to check that the following vector \cite{shwa}
\begin{eqnarray}
|z\rangle\!\rangle &=& \int_{-\infty}^{+\infty}\frac{dx}{\sqrt{\pi}}
e^{2ix\hbox{\scriptsize Im}z}|x\rangle_{0}\otimes|\hbox{Re}z-x
\rangle_{0}\nonumber  \\
&=&\int_{-\infty}^{+\infty}\frac{dy}{\sqrt{\pi}}
e^{-2iy\hbox{\scriptsize Re}z}|y+\hbox{Im}z\rangle_{\pi/2}
\otimes |y\rangle_{\pi /2}\;\label{zeta}
\end{eqnarray}
is eigenvector of $\hat Z$ with complex eigenvalue $z$. In Eq.~
(\ref{zeta}) $|\psi \rangle \otimes |\varphi \rangle $ denotes a vector 
in the two-mode Hilbert space $\cal H=\cal H_{a}\otimes\cal H_{b}$, and 
$|x\rangle_{\phi}$ represents an eigenvector of the quadrature 
$\hat{X}_{\phi}=\frac{1}{2}(c^{\dag}e^{i\phi}+\hbox{h.c.})$ of 
the pertaining mode $c=a,b$. The notation $|\ \rangle\!\rangle$ remembers 
that the state is a two-mode one. The set $\{|z\rangle\!\rangle\}$ is 
complete orthonormal for $\cal H$, with scalar product:
\begin{eqnarray}
\langle\!\langle z|z'\rangle\!\rangle =\delta^{(2)}(z-z')\equiv 
\delta (\hbox{Re}z-\hbox{Re}z')\,\delta (\hbox{Im}z-\hbox{Im}z')\,.\!\!\!
\label{delta}
\end{eqnarray}
In the number representation the vector (\ref{zeta}) reads as follows
\begin{eqnarray}
|z\rangle\!\rangle = e^{i\hbox{\scriptsize Re}z\hbox{\scriptsize Im}z}
\sum_{n,m=0}^\infty \hbox{c}_{n,m}(z,\overline{z})|n\rangle\otimes|m\rangle\;,
\label{eigen}
\end{eqnarray}
with
\begin{eqnarray}
&&\hbox{c}_{n,n+\lambda}(z,\overline{z})=\overline{\hbox{c}}_{n+\lambda,n}
(z,\overline{z})=\nonumber\\
&&=\frac{(-)^{n}}{\sqrt{\pi}}\sqrt{\frac{n!}{(n+\lambda)!}}
\,\overline{z}^{\lambda}\,\hbox{L}_{n}^{\lambda}(|z|^{2})\,\exp
\left(-\frac{1}{2}|z|^{2}\right) 
\label{cn}\;.
\end{eqnarray}
Eq. (\ref{cn}) is obtained from Eq. (\ref{zeta}) using the number 
representation of the quadrature 
\begin{eqnarray}
{}_{\phi }\langle x|n\rangle = \left( \frac{2}{\pi}\right)^{1/4}\frac
{e^{in\phi}}{\sqrt{2^{n}n!}}e^{-x^2}\hbox{H}_{n}(\sqrt{2}\,x)\;,\label{herm}
\end{eqnarray}
along with the following identity between Hermite and Laguerre 
polynomials
\begin{eqnarray}
&&\int_{-\infty}^{+\infty}\frac{dx}{\sqrt{\pi}}e^{-x^{2}}
\hbox{H}_{n}(x+y)\hbox{H}_{n+\lambda}(x+t)\nonumber\\
&&=2^{n+\lambda }\,n!\,\hbox{L}_{n}^{\lambda}(-2yt)\,t^{\lambda}\;.\label{lag}
\end{eqnarray}
The Dirac-normalized states $|z\rangle\!\rangle$ have infinite total 
number of photons, and we seek physically realizable states 
approaching $|z\rangle\!\rangle$ for infinite photon numbers. The 
eigenstate corresponding to zero eigenvalue is given by:
\begin{eqnarray}
|0\rangle\!\rangle=\frac{1}{\sqrt{\pi}}\sum_{n=0}^{\infty}(-)^{n}
|n\rangle\otimes|n\rangle\;.\label{twin}                              
\end{eqnarray}
This is just the ``twin-beams'' at the output of a phase-insensitive 
amplifier (PIA) in the limit of infinite gain \cite{mauro1}. One has
\begin{eqnarray}
|0\rangle\!\rangle=\lim_{\lambda\rightarrow 1^{-}}
|0\rangle\!\rangle_{\lambda}\;,
\end{eqnarray}
with
\begin{eqnarray}
|0\rangle\!\rangle_{\lambda}&=&(1-\lambda ^{2})^{1/2}  
\sum_{n=0}^{\infty}(-\lambda)^{n}|n\rangle\otimes|n\rangle= \nonumber \\
&=&\exp [\hbox{tanh}^{-1}\,\lambda\,(ab-a^{\dag}b^{\dag})]\,
|0\rangle\otimes|0\rangle\;.\label{zero}
\end{eqnarray}
In the parametric approximation of infinite classical (undepleted) 
pump the modes $a$ and $b$ are identified with a couple of signal 
and idler modes of the amplifier (the gain is $(1-\lambda ^{2})^{-1}$). 
Apart from an irrelevant phase factor, the eigenstate $|z\rangle\!\rangle$ 
can be generated by $|0\rangle\!\rangle$ upon 
displacing either $a$ or $b$. Displacing the mode $a$ we have
\begin{eqnarray}
|z\rangle\!\rangle=e^{i\phi_{z}}e^{za^{\dag}-\overline{z}a}
|0\rangle\!\rangle\;.\label{disp}
\end{eqnarray}
The physical (normalizable) state $|z\rangle\!\rangle_{\lambda}$ 
approaching $|z\rangle\!\rangle $ for infinite gain is obtained in the 
same way
\begin{eqnarray}
|z\rangle\!\rangle_{\lambda}=e^{i\phi_{z}}e^{za^{\dag}-\overline{z}a}
(1-\lambda ^{2})^{1/2}  
\sum_{n=0}^{\infty}(-\lambda)^{n}|n\rangle\otimes|n\rangle
\;.\label{zetal}
\end{eqnarray}
The displacement in Eq. (\ref{zetal}) can be achieved by combining the 
``twin-beams'' $|0\rangle\!\rangle_\lambda $ with a strong coherent local 
oscillator $|\beta\rangle \ (\beta \rightarrow \infty)$ in a beam splitter 
with a transmissivity $\tau \rightarrow 1$, such that 
$|\beta|\sqrt{1-\tau}=|z|$ (the local oscillator is at the frequency 
of the signal mode $a$). 
\par The experimental set-up to generate the state (\ref{zetal}) is sketched 
in Fig.~1. 
The state (\ref{zetal}) has average number of photons
\begin{eqnarray}
\overline{n}={}_{\lambda}\langle\!\langle z|a^{\dag}a+b^{\dag}b
|z\rangle\!\rangle {}_{\lambda}=|z|^{2}+\frac{2\lambda^2}{1-\lambda^2}
\;.\label{num}
\end{eqnarray}
The state (\ref{zetal}) is now impinged into a heterodyne 
detector with signal mode $a$ and image-band mode $b$. The probability 
density of getting the value $z$ for the output photocurrent $\hat Z$ 
with the field in the state $|w\rangle\!\rangle_{\lambda}$ is given by
\begin{eqnarray}
|\langle\!\langle z|w\rangle\!\rangle_{\lambda}|^{2}  
&=&(1-\lambda ^{2})\left|\sum_{n=0}^{\infty}(-\lambda)^{n}
\,\hbox{c}_{n,n}(z-w,\overline{z-w})\right|^{2}\nonumber\\            
&=&\frac{1-\lambda^{2}}{\pi}\exp{(-|z-w|^{2})}
\left|\sum_{n=0}^{\infty}\lambda^{n}\hbox{L}_{n}(|z-w|^{2})\right|^2
\nonumber \\
&=&\frac{1}{\pi\Delta_{\lambda}^{2}}\exp\left( {-\frac{|z-w|^{2}}
{\Delta_{\lambda}^{2}}}\right)    \;\label{prob}
\end{eqnarray}
where
\begin{eqnarray}
\Delta_{\lambda}^{2}=\frac{1-\lambda }{1+\lambda }\;.\label{D}
\end{eqnarray}
In the limit $\lambda \rightarrow 1^-$ one has that 
$|\langle\!\langle z|w\rangle\!\rangle_{\lambda}|^{2}\rightarrow \delta
^{(2)}(z-w)$, confirming that the state $|w\rangle\!\rangle_{\lambda}$ 
approaches the eigenstate $|w\rangle\!\rangle$ of the current $\hat Z$\@.
\par The detection of the phase $\hat{\phi}=\arg(\hat{Z})$ is 
described by the marginal probability density of (\ref{prob}), namely
\begin{eqnarray}
&&p(\phi)=\frac{1}{\pi \Delta_{\lambda}^{2}}\int_{0}^{+\infty}dr
\,r\,\exp\left( -{|re^{i\phi}-|w|e^{i\theta}|^2
\over \Delta_{\lambda}^{2}}\right) \nonumber \\
&&={1\over 2\pi}e^{-{|w|^{2}\over\Delta_{\lambda}^{2} }}
+{|w|\over\pi\Delta_{\lambda} }\cos(\phi-\theta)
\nonumber \\&&\times{\sqrt{\pi}\over 2}
\,\left[1+\hbox{erf}\left({|w|\cos(\phi -\theta)\over\Delta_{\lambda} }\right) 
\right]\,e^{-{|w|^{2}\over \Delta_{\lambda}^{2}}\sin ^{2}(\phi -\theta)} 
\;,\label{pfi}
\end{eqnarray}
where $\theta =\arg(w)$, and $\hbox{erf}(x)$ denotes the error 
function $\hbox{erf}(x)={2\over\sqrt{\pi}}\int_{0}^{x}dt\,e^{-t^2}$. 
Notice that the probability density (\ref{pfi}) is just the Born rule 
for the self-adjoint operator $\hat{\phi}=\arg(\hat{Z})=-{i\over2}
\log(\hat Z/\hat{Z}^{\dag})$: this is well defined on the 
Hilbert space ${\cal H}_{0}^{\bot}$, orthogonal complement in $\cal H$ 
of the space ${\cal H}_{0}$ spanned by vector $|0\rangle\!\rangle$ in 
Eq. (\ref{twin}) \cite{hrad}. The integral over $r$ in Eq. (\ref{pfi}) 
just sums up degeneracies of eigenvectors (\ref{eigen}): the zero-eigenvalue 
vector is not degenerate, and gives a zero-measure contribution to the 
integral. The first Gaussian term in the last side of Eq.~(\ref{pfi}) 
gives a uniform phase probability distribution for the ``twin-beams'' 
input state $|0\rangle\!\rangle_\lambda $. 
\par For $\Delta_{\lambda}\ll |w|$ Eq. (\ref{pfi}) approaches the Gaussian form
\begin{eqnarray}
p(\phi)\simeq\frac{|w|}{\sqrt{\pi}\Delta_{\lambda}}\exp\left[ -{|w|^2
\over\Delta_{\lambda}^2}(\phi -\theta)^2\right] \;\label{Ga}
\end{eqnarray}
corresponding to the r.m.s. phase sensitivity \cite{cramer}
\begin{eqnarray}
\delta\phi=\langle\Delta\phi^{2}\rangle^{1/2}=
{1\over\sqrt 2}{\Delta_{\lambda}\over|w|}\;.
\end{eqnarray}
In the limit of infinite gain at the PIA $(\lambda\rightarrow 1^-)$ one 
has $\Delta_{\lambda}^{2}\simeq{1\over2}(1-\lambda)$ and $\bar n\simeq
|w|^{2}+(1-\lambda)^{-1}$. [Notice that the classical approximation for the 
local oscillator at the beam splitter requires that its intensity $|\beta |^2$ 
must be much greater than the input photon number 
$\simeq (1-\lambda )^{-1}$ of the ``twin beams''.]
Optimizing $\delta \phi$ versus $|w|$ at 
fixed $\bar n$ one obtains the sensitivity 
\begin{eqnarray}
\delta\phi\simeq{1\over\bar n}\;\label{1sun}
\end{eqnarray}
for $|w|^2=(1-\lambda)^{-1}$, namely for signal photons equal to the 
``twin-beams'' photons. The sensitivity (\ref{1sun}) obeys the same 
power-law as the ideal sensitivity for one-mode phase detection 
(actually it is improved by a constant factor equal to 1.36: see Ref. 
\cite{rev1}).
\par The ideal phase sensitivity (\ref{1sun}) has been derived with the 
hypothesis of unit efficiency at the heterodyne photodetector. It is 
easy to show that for nonunit quantum efficiency 
(independent on frequency in the range between signal and image-band 
modes) Eq. (\ref{D}) becomes
\begin{eqnarray}
\Delta_{\lambda}^2 \rightarrow \Delta_{\lambda}^2(\eta)=
\Delta_{\lambda}^2+{1-\eta\over\eta}\;.
\end{eqnarray}
Then, it is clear that the result (\ref{1sun}) holds only in the limit 
$1-\eta\ll|w|^{-2}$, whereas in the opposite situation $1-\eta\gg|w|^{-2}$ 
one obtains the usual shot noise $\delta\phi =\sqrt{(1-\eta)/2\bar n}$.
\par In conclusion, we have presented a feasible scheme to detect a 
two-mode phase of the field, approaching the eigenstates of the 
heterodyne current $\hat Z$. The state of the field is obtained by 
means of a high gain PIA followed by a high transmissivity beam splitter 
with strong local oscillator at the signal frequency. The ideal r.m.s. 
sensitivity $\delta\phi=1/\bar n$ is achieved for large photon numbers 
$\bar n\gg1$ and for signal photons $|w|^{2}=\bar {n}/2$. The gain of 
the PIA (parametrically ideal) is tuned to the value 
$g={1\over 4}\bar n$, and the quantum efficiency at the photodetector 
must be very good, namely 
$1-\eta\ll 2/{\bar n}$.
\par Hence, the two-mode phase detection could be experimentally achieved, 
but the technical requirements are strict: two local oscillators plus a 
classical pump at the PIA (all of them coherent and at different 
frequencies); linear amplification for high gains, 
with the pump still undepleted; very good quantum efficiency. This 
shows how technical difficulties can rise when going from one-mode 
to two-mode phase detection. 

\medskip 
One of us (G. M. D'Ariano) acknowledges stimulating discussions with 
P. Kumar and H. P. Yuen.

\end{multicols}
\begin{figure}[htb]
\begin{center}
\setlength{\unitlength}{0.7cm}
      \begin{picture}(23,7)(-2,-2.5)
      \thicklines
      \drawline(-2,1)(0,1)
      \drawline(-2,2)(0,2)
      \put(-3.3,1.8){$|0\rangle$}
      \put(-3.3,0.8){$|0\rangle$}
      \put(0,0){\framebox(4.5,3){{\footnotesize 
      $\matrix{\hbox{\large PIA}\cr
      {\hbox{gain}}\,(1-\lambda^{2})^{-1/2}} $}}}       
      \drawline(4.5,1)(6,1)
      \drawline(4.5,2)(6,2)
      \drawline(-1.5,-1.5)(0,0)
      \drawline(4.5,3)(6,4.5)
      \put(5.2,3.7){\vector(1,1){0}}
      \put(-0.8,-0.8){\vector(1,1){0}}
      \put(-1.5,-2){\large{strong pump}}
      \put(6,2.5){\large{``twin-beams''}}
      \put(7.5,1.3){\large{$|0\rangle\!\rangle_{\lambda}$}}   
      \drawline(10,1)(16.5,1)
      \drawline(10,2)(16.5,2)
      \drawline(14.1,-1)(14.1,2)
      \put(14.1,0){\vector(0,1){0}}
      \drawline(15,3.05)(12.2,0.25)  
      \drawline(15.1,2.95)(12.3,0.15)  
      \drawline(15.1,2.95)(15,3.05)  
      \drawline(12.3,0.15)(12.2,0.25)
      \put(14.8,-0.5){$|w\rangle$}
      \put(11.5,-2){$\sqrt{1-\theta}|w|=|z|$}
      \put(12,-0.5){$\theta\rightarrow 1$}   
      \put(17.4,1.3){\large{$|z\rangle\!\rangle_{\lambda}$}}   
      \put(18.9,1.4){$\rightarrow $}   
      \put(21.2,1.6){\makebox(0,0){$\matrix{\hbox{to the}\cr
      \hbox{heterodyne}\cr \hbox{detector}} $}}       
      \end{picture}
\end{center}
\caption{Outline of the experimental setup to generate two-mode phase states
approaching heterodyne eigenstates. The PIA produces the ``twin-beams'' 
in Eq. (\protect{\ref{zero}}) and the beam splitter achieves the displacement 
(\protect{\ref{zetal}}) [see text].}
\label{f:outline}
\end{figure}
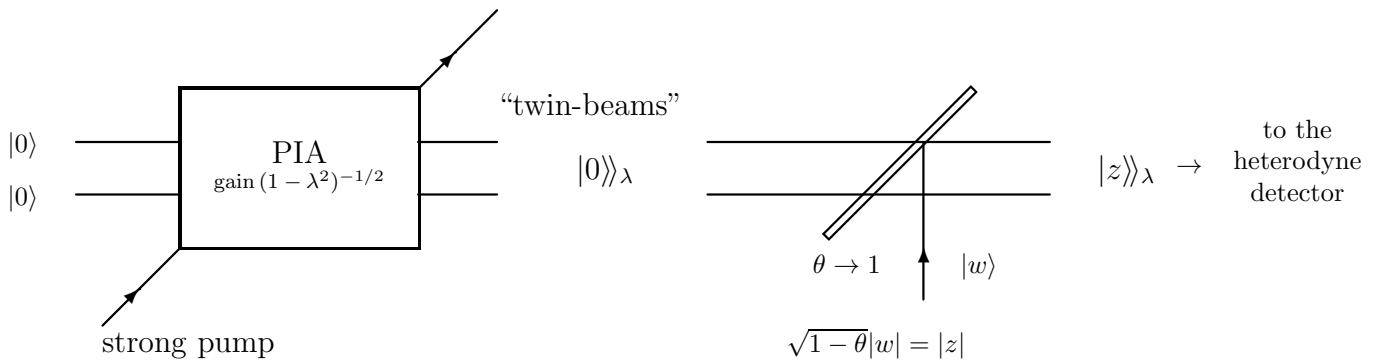                                                                   
\end{document}